# Competing Charge/Spin-Stripe and Correlated Metal Phases in Trilayer Nickelates $(Pr_{1-x}La_x)_4Ni_3O_8$

Xinglong Chen,[1,*] Hong Zheng,[1] Daniel Phelan,[1] Hao Zheng,[1] Saul Lapidus,[2] Matthew J. Krogstad,[2] Raymond Osborn,[1] Stephan Rosenkranz,[1] and J. F. Mitchell[1,*]

[1]Materials Science Division, Argonne National Laboratory, Lemont, Illinois 60439, United States

[2]Advanced Photon Source, Argonne National Laboratory, Lemont, Illinois 60439, United States

**ABSTRACT:** Low-valent nickelates $R_{n+1}Ni_nO_{2n+2}$ ($R$ = rare earth) containing $Ni^{1+}$ ($d^9$) with a quasi-two-dimensional (quasi-2D) square planar coordination geometry possess structural and electronic properties that are similar to those of high-$T_c$ cuprates, including superconductivity itself in the doped infinite layer ($n = \infty$) $RNiO_2$ system. Within this $R_{n+1}Ni_nO_{2n+2}$ nickelate family, the crystallographic isomorphs $Pr_4Ni_3O_8$ and $La_4Ni_3O_8$ exhibit singularly different ground states: $Pr_4Ni_3O_8$ is metallic and $La_4Ni_3O_8$ is a charge- and spin-stripe ordered insulator. To explore and understand the ground state evolution from metallic $Pr_4Ni_3O_8$ to stripe-ordered $La_4Ni_3O_8$ in the $R_4Ni_3O_8$ family, we have grown a series of isovalent-substituted single crystals $(Pr_{1-x}La_x)_4Ni_3O_8$. Combining thermodynamic, transport, magnetic, and synchrotron X-ray single crystal diffraction measurements, we reveal a transition between metallic and stripe-insulator phase regions, with a putative quantum phase transition at $x \approx 0.4$. We propose two possible models for $(Pr_{1-x}La_x)_4Ni_3O_8$: an electronically inhomogeneous system that could serve as a candidate for exploring quantum Griffiths phase physics and a homogeneous system with a putative quantum critical point at the phase boundary.

**INTRODUCTION**

Nickel oxides occupy a privileged niche in the physics and chemistry of transition metal oxides, a recognition they have earned largely because of John Goodenough's pioneering insights into the magnetic and electronic and structural properties of the $R$NiO$_3$ ($R$ = rare earth) family of perovskite nickelates.[1—6] Today, Goodenough's legacy of past decades still inspires widespread, intense study of not only these archetypal perovskites but also other related nickelates where competition between localized and itinerant electrons underpins the physics.

In this vein, mixed-valent layered nickelates $R_{n+1}$Ni$_n$O$_{2n+2}$ that contain Ni$^{1+}$ ($d^9$) in a quasi-two-dimensional (quasi-2D) square planar geometry have recently been in focus as platforms for exploring the physics of strongly correlated electron systems and as model systems for understanding high $T_c$ superconductivity. Indeed, after a three-decade search, superconductivity was discovered in 2019 in the aliovalent ion-doped infinite layer ($n = \infty$) nickelate thin film (Nd,Sr)NiO$_2$,[7] and subsequently in its analogs (Pr,Sr)NiO$_2$,[8] and (La,Sr)NiO$_2$.[9] More recently superconductivity has also been reported in a self-doped quintuple-layer ($n = 5$) Nd$_6$Ni$_5$O$_{12}$ thin film.[10] The community is now actively pursuing a deeper understanding of these thin film nickelates, including how much they share with cuprates,[11,12] and Bernadini *et al*. have recently surveyed issues that influence thin film nickelate phenomenology.[13] To date, however, no *bulk* superconducting infinite layer superconductors have been reported.

Although not superconducting, the $n = 3$ member of $R_{n+1}$Ni$_n$O$_{2n+2}$, trilayer $R_4$Ni$_3$O$_8$, ($R$ = La, Pr, Nd, Sm),[14—18] which possesses an average Ni valence of 1.33+ ($d^{8.67}$), crystallizes in the tetragonal $I4/mmm$ space group and features stacking of $R$O$_2$ fluorite layers and triple infinite NiO$_2$ square-planar layers (Figure 1a). In a series of studies,[19—23] we have shown that bulk crystals of $R_4$Ni$_3$O$_8$ possess many of the 'ingredients' thought important to cuprate superconductivity and may only suffer from lying nominally in the overdoped regime of the high $T_c$ electronic phase diagram. In particular, we have reported that when $R = $ Pr, crystals are metallic[20] while for $R = $ La, a charge- and spin-stripe order emerge as the ground state.[19,21] Thus, it is expected that a $T = 0$ quantum phase transition (QPT) between a correlated metal and a charge- and spin-stripe ordered insulator will be found along the $(Pr_{1-x}La_x)_4Ni_3O_8$ composition line. Depending on the nature of the phase evolution with $x$—heterogeneous and discontinuous or homogeneous and continuous—such a putative QPT could correspond respectively to potential host for quantum Griffiths phase physics or to a quantum critical point.[24-26]

In this context, we set out to explore and understand the evolution of the competing ground states from metallic $Pr_4Ni_3O_8$ to insulating $La_4Ni_3O_8$ in the $R_4Ni_3O_8$ family. We report here the first synthesis of $(Pr_{1-x}La_x)_4Ni_3O_8$ single crystals and a detailed experimental investigation of the $(Pr_{1-x}La_x)_4Ni_3O_8$ solid solution using these crystals. A combination of magnetic, transport, thermodynamic and scattering measurements were used to establish the $(x, T)$ electronic and magnetic phase diagram of this system and to locate the QPT at $x \approx 0.4$. We find that the stripe phase is remarkably insensitive to composition in its field of stability, remaining commensurate with its in-plane propagation vector unchanged from the $\boldsymbol{q} = (1/3, 1/3)$ found in the $La_4Ni_3O_8$ parent phase.[19] Furthermore, no observable change in the spatial coherence of the stripes is found as the phase boundary is approached. To explain these observations, we propose two possible models for the behavior: (1) an inhomogeneous model in which growth of Pr-rich regions progressively consumes a La-rich matrix of charge- and spin-stripe order, a process that would make possible a quantum Griffiths phase scenario, and (2) a homogeneous model in which the amplitude of the spin- and charge-order various continuously toward a quantum critical point.

## EXPERIMENTAL SECTION

**Syntheses and Crystal Growth.** Precursor powders of $(Pr_{1-x}La_x)_4Ni_3O_{10}$ ($x = 0.1, 0.3, 0.4, 0.5, 0.7$ and 0.9) for single crystal growth were synthesized by a standard solid-state reaction using stoichiometric amounts of $Pr_6O_{11}$ (Alfa Aesar, 99.99%), $La_2O_3$ (Alfa Aesar, 99.999%) and NiO (Strem Chemicals, 99.99%) at 1050 °C. The powders were hydrostatically pressed into polycrystalline rods and sintered for 24–48 h at 1100–1400 °C to make dense rods. Single crystals of $(Pr_{1-x}La_x)_4Ni_3O_{10}$ ($x = 0.1, 0.3, 0.4, 0.5, 0.7$ and 0.9) were successfully grown from the sintered rods using high-pressure floating-zone furnace (HKZ-1; SciDre GmbH) under various oxygen pressures ($pO_2$) in the range of 25-135 bar. Single crystals of $(Pr_{1-x}La_x)_4Ni_3O_8$ ($x = 0.1, 0.3, 0.4, 0.5, 0.7$ and 0.9) were obtained by using topotactic reduction (4 mol% $H_2/N_2$ gas; 350 °C; 5 d) of $(Pr_{1-x}La_x)_4Ni_3O_{10}$ specimens separated from the grown boules.

**Powder X-ray Diffraction.** Phase identification of pulverized crystals of $(Pr_{1-x}La_x)_4Ni_3O_8$ was performed by powder X-ray diffraction (PXRD) using a PANalytical X'Pert PRO diffractometer equipped with Cu K$\alpha$ radiation ($\lambda$ = 1.5418 Å). Room temperature lattice constants were determined from high-resolution powder synchrotron X-ray diffraction measurements carried out at Beamline 11-BM-B at the Advanced Photon Source (APS) at Argonne National Laboratory with l = 0.458192 Å in the range of $0.5° \leq 2\theta \leq 30°$, with a step size of 0.001° and counting time of 0.1 s. Data were analyzed by Pawley fitting of the profiles using Bruker AXS Topas 6.0.

**Single Crystal X-ray Diffraction:** Single crystal X-ray diffraction data of $(Pr_{1-x}La_x)_4Ni_3O_8$ ($x = 0.5, 0.7$ and 0.9) were collected on Sector 6-ID-D at the Advanced Photon Source (APS) at Argonne National Laboratory using an incident energy of 87.1 keV (wavelength of 0.142 Å). A three-dimensional volume of intensity was constructed via a series of 3700 images, each collected during a total rotation of 370°, using a Dectris Pilatus 2M containing a CdTe detecting layer of thickness 1 mm. Transformation from detector space to reciprocal space was performed using CCTW (Crystal Coordinate Transformation Workflow).[27]

**Thermal Properties.** The heat capacity ($C_p$) of $(Pr_{1-x}La_x)_4Ni_3O_8$ single crystals was measured in the temperature range 1.8–300 K using the relaxation method on a Quantum Design Dynacool Physical Properties Measurement System (PPMS). Apiezon-N vacuum grease was used to fix the crystals onto the sample holder platform.

**Magnetic Properties.** Magnetic susceptibility measurements were performed using a Quantum Design MPMS3 SQUID Magnetometer. Single crystals were fixed to a quartz holder using GE vanish. For each composition, data with the field applied in the ***ab***-plane and along the ***c***-axis were collected, in the range 1.8–300 K under a magnetic field of 5 T under both zero-field cool (ZFC) and field cool (FC) conditions. For $(Pr_{1-x}La_x)_4Ni_3O_8$ ($x = 0.4$ and 0.5), ZFC and FC data were also collected along the ***c***-axis under various magnetic fields (0.01 T, 0.1 T, 0.3 T, 1 T, 2T, 5 T and 7 T) between 1.8 and 300 K. Isothermal field-dependent magnetization was measured on $(Pr_{1-x}La_x)_4Ni_3O_8$ ($x = 0.4$ and 0.5) single crystals with magnetic field applied parallel to the ***ab***-plane and along the ***c***-axis in the field range of 4–7 T at various temperatures between 1.8 K and 300 K. Isothermal field-dependent magnetization measurements in the field range of ±7 T were also measured on $(Pr_{1-x}La_x)_4Ni_3O_8$ ($x = 0.5$) at 1.8 K with magnetic field parallel to the ***ab***-plane and along the ***c***-axis.

**Electrical Properties.** Electrical resistivity measurements were carried out in the *ab*-plane of the single crystal samples using the standard four-probe technique under zero magnetic field on a Quantum Design PPMS in the temperature range 1.8–300 K.

**RESULTS**

**Material Synthesis.** Crystalline boules of $(Pr_{1-x}La_x)_4Ni_3O_{10}$ ($x = 0.1, 0.3, 0.4, 0.5, 0.7$ and $0.9$) were grown in a floating zone furnace at various $pO_2$ ranging from 25 to 135 bar, depending on the composition (Table 1). Black single crystals of $(Pr_{1-x}La_x)_4Ni_3O_8$ ($x = 0.1, 0.3, 0.4, 0.5, 0.7$ and $0.9$) with millimeter-sizes were obtained by reducing corresponding $(Pr_{1-x}La_x)_4Ni_3O_{10}$ specimens cleaved from the boules. The obtained $(Pr_{1-x}La_x)_4Ni_3O_8$ single crystals are fragile, and their sizes (typically $\sim 0.5$–$2$ mm on a side) are much smaller than those of the parent $(Pr_{1-x}La_x)_4Ni_3O_{10}$ due to the apparently unavoidable cracking that occurs during the topochemical reduction process, during which the crystals undergo a large anisotropic volume change. Even so, we noticed that selecting well-formed $(Pr_{1-x}La_x)_4Ni_3O_{10}$ single crystals for reduction is helpful to obtain low crack-density single crystals of the $(Pr_{1-x}La_x)_4Ni_3O_8$ series.

**Table 1.** Crystal growth parameters, room temperature unit cell parameters, and transition temperatures of $(Pr_{1-x}La_x)_4Ni_3O_8$. Numbers in parentheses are estimated standard deviations. $T_{CO}$ defined as intersection of tangent lines to the curve. $T_N$ defined as inflection point.

| | $x$ in $(Pr_{1-x}La_x)_4Ni_3O_8$ | | | | | |
| --- | --- | --- | --- | --- | --- | --- |
| | 0.1 | 0.3 | 0.4 | 0.5 | 0.7 | 0.9 |
| Growth $pO_2$ (bar) | 135 | 130 | 110 | 100 | 60 | 25 |
| Growth Rate (mm/hr) | 5 | 3–5 | 3–5 | 3–5 | 4–5 | 5 |
| $a$ (Å) | 3.93444(2) | 3.94252(1) | | 3.94995(4) | 3.9579(3) | 3.96521(4) |
| $c$ (Å) | 25.5404(2) | 25.6704(1) | | 25.7925(3) | 25.9204(3) | 26.0304(3) |
| $V$ (Å³) | 395.36(1) | 399.01(1) | | 402.42(1) | 406.04(1) | 409.27(1) |
| $T_{CO,}$ cooling (K) | | | | 54 | 84 | 99 |
| $T_{CO,}$ warming (K) | | | | 64 | 88 | 101 |
| $T_N$, warming (K) | | | | 56 | 84 | 97 |

**Crystal Structure.** The crystal structure of $(Pr_{1-x}La_x)_4Ni_3O_8$ is shown in Figure 1a. It is characterized by square-planar $NiO_2$ sheets condensed into a trilayer unit, with these trilayer units separated by fluorite-like blocks of $(Pr,La)O_2$ along the *c*-axis. In this crystal structure, there are two distinct rare earth sites, one in the Ni-O trilayer and one in the fluorite layer. In principle, the La and Pr could be distributed nonuniformly across these sites; however, we have not explored this possibility and assume a uniform average occupation.

**PXRD Refinement.** A typical Pawley refinement of the room temperature synchrotron X-ray powder diffraction data collected on crushed single crystals and based on a single *I4/mmm* phase is shown in Figure 1b, and the results of such fits across the series are collected in Table 1. As shown in Figure 1c, the 298 K unit cell volume of $(Pr_{1-x}La_x)_4Ni_3O_8$ expands linearly with the increase of La content following a Vegard's law behavior. The volume of $x = 0$ and $x = 1$ specimens determined from our earlier work[19,20] fit on the extrapolation of this line. Together, these results establish successful growth of the entire solid solution joining the Pr- and La-endmembers.

**Transport and Magnetic Behavior.** Transport measurements establish that the $T = 0$ phase transition occurs near $x = 0.4$. The ***ab***-plane electrical resistivity normalized to 300 K is shown in Figure 2a. Specimens in the range $x = 0.1-0.3$ exhibit a metallic temperature dependence ($dR/dT > 0$) in the entire temperature range, and a finite resistance as $T \rightarrow 0$ K. This behavior is analogous to that of the Pr ($x = 0$) endmember described in Ref. 20. The rather poor RRR $\approx 10$ for these samples likely reflects a substantial concentration of defects in these reduced crystals, such as the microcracks mentioned earlier. Nonetheless, these samples are intrinsically metallic. In contrast, samples with $x \geq 0.5$ show a dramatic upturn with hysteresis in resistance upon cooling below a metal-insulator transition (MIT) temperature, $T_{MIT}$, indicating a first-order phase transition into a highly localized electronic state. The hysteresis narrows as $x$ approaches 1.0, correlating with reduced Pr/La site mixing. The more than three-decade increase in resistance for $T < T_{MIT}$ is a lower bound, as the measured resistance exceeded the upper limit of our instrumentation. We have previously identified a similar transition in the La endmember ($x = 1$), where it has been attributed to the formation of charge and spin stripes.[19] As we show below, the origin of the MIT in the samples with $x > 0.4$ can also be attributed to the onset of charge stripes.

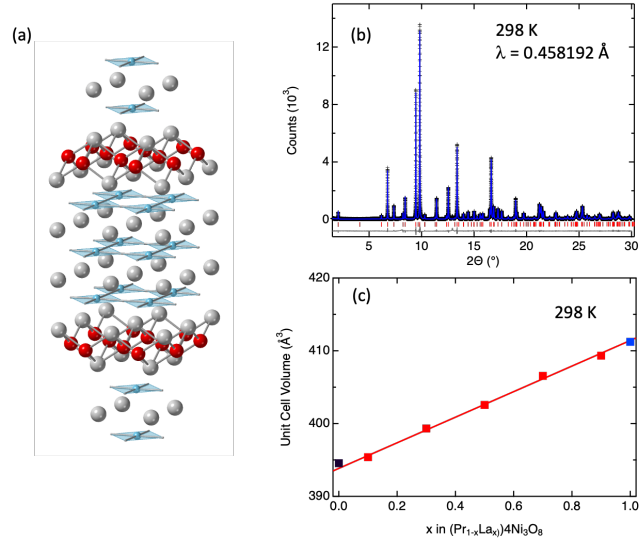

**Figure 1.** (a) Crystal structure of (Pr,La)$_4$Ni$_3$O$_8$. Light blue spheres, Ni; gray spheres, Pr/La; red spheres, O. (b) Pawley refinement of synchrotron X-ray powder data. Crosses: data; solid blue line: model; solid gray line: difference; vertical lines mark position of peaks expected for *I4/mmm*. (c) Variation of unit cell volume with La content in (Pr$_{1−x}$La$_x$)$_4$Ni$_3$O$_8$. Red squares and solid fit line are from this work; blue and black squares from Ref. 19 and 20, respectively. Error bars (esd) are smaller than the symbols.

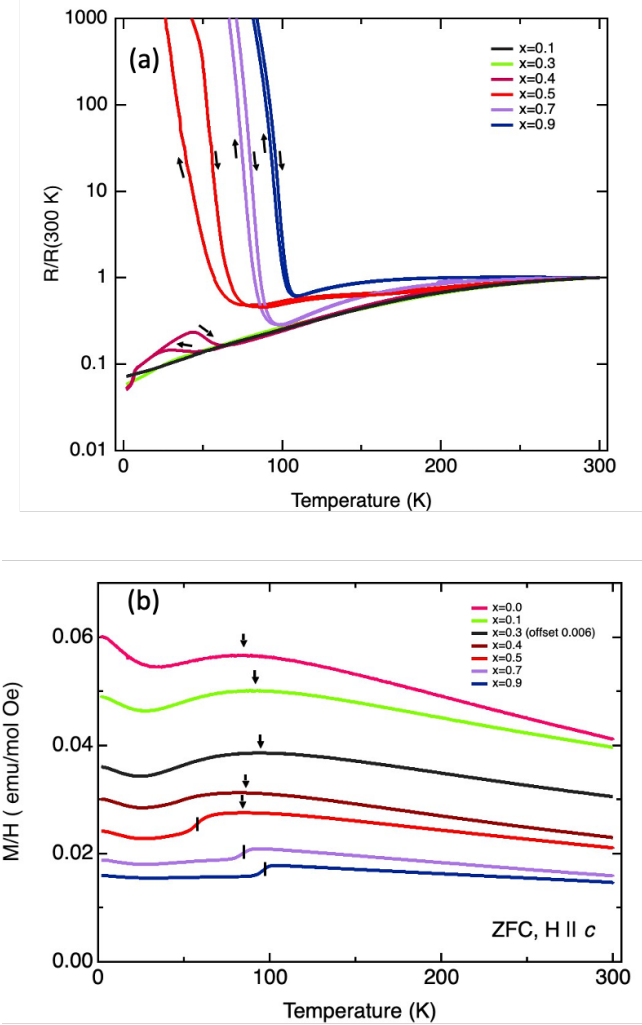

**Figure 2**. (a) Normalized ***ab***-plane electrical resistance of $(Pr_{1-x}La_x)_4Ni_3O_8$ single crystals vs. temperature, measured on cooling and warming. The data for $x = 0.3$ has been offset vertically by +0.006 emu/mol Oe for clarity. (b) ***c***-axis $M/H$ measured in $\mu_oH$ = 5 T after zero-field cooling to 1.8 K. Vertical lines mark the inflection point defining the magnetic transition, while arrows track the position of a broad maximum.

The $x = 0.4$ sample shows a weak hysteresis superimposed on a metallic background. It is possible that this reflects some intermediate state between the metal and charge-stripe insulator. However, it is more likely that this small effect is a consequence of inevitable concentration fluctuations of Pr and La in a finite sized specimen. That is, the bulk of the crystal is metallic, but small regions with x > 0.4, which are insulating, lie in the current path. The large (>10⁴) contrast in resistance between metal and insulator leads to the appearance of this small (< factor of 2) effect in the bulk transport. We have described similar behavior in bilayer manganites near the boundary between charge-stripe ordered and metallic phases.[28] Thus, based on transport, the $T = 0$ K transition between metal and stripe-insulator lies at $x \approx 0.4$.

Magnetic data support the finding of the phase boundary at $x \approx 0.4$. A significant anisotropy between $\textbf{\textit{ab}}$-plane and $\textbf{\textit{c}}$-axis susceptibility emerges below 300 K, growing to nearly a factor of four at 1.8 K (See Figure S4b in the SI). Figure 2b shows the $\textbf{\textit{c}}$-axis $M/H$ vs. temperature for several compositions of $(Pr_{1-x}La_x)_4Ni_3O_8$. All specimens with $x \leq 0.5$ show a broad maximum on cooling between 80 and 100 K (arrows), while those with $x \geq 0.5$ evidence an abrupt kink downward (vertical marks). This kink is characteristic of the formation of spin stripes concomitant with charge stripes, as we have shown for $La_4Ni_3O_8$ ($x = 1$) using single crystal neutron diffraction.[21] Additionally, a low-temperature minimum is found near 30 K, which is nearly independent of composition, followed by an upturn that becomes more pronounced in the Pr-rich region. In fact, as shown in the SI, the size of this upturn scales linearly with $(1-x)$, vanishing for the La endmember. We thus attribute this feature to Pr-localized spins rather than to the Ni sublattice, which participates in the spin stripes. Other authors have reported a similar upturn in powder samples.[29] However, single crystals reveal that the upturn is absent in the $\textbf{\textit{ab}}$-plane magnetization (see Figure S1), an anisotropy that is likely induced by the crystal field acting on the Pr spins (more detailed discussion can be found in the SI).

**Heat Capacity.** The heat capacity of crystals with all compositions is shown in Figure 3a as $Cp/T$ vs $T$. In the main panel, specimens of $x = 0.5, 0.7$, and $0.9$ show clear anomalies associated with the transition to the charge- and spin-stripe phase. The inset shows that samples with $x \leq 0.4$ follow a smooth temperature variation with no features visible, consistent with these samples remaining metallic to 1.8 K. The entropy associated with the transition can be extracted by integrating the area under the peak in these data. In the absence of a nonmagnetic analog, we have phenomenologically fit a polynomial to the data on either side of the peak to serve as a background. The extracted peaks are shown in the main panel of Figure 3b along with the integral, $\Delta S_{mag}$, which we attribute to the condensation of entropy in the magnetic channel as spin-stripes form. Included is the $x = 1$ endmember, which we previously showed using this same background subtraction process to have a $\Delta S_{mag} \approx 6$ J/mol-K, in good agreement with that found in stripe phases of single layer nickelates, $La_{2-x}Sr_xNiO_4$.[30] As Pr substitutes for La, the peak in $C_p/T$ is quickly suppressed with a corresponding decrease in $\Delta S_{mag}$. The inset to Figure 3b shows how $\Delta S_{mag}$ (scaled to a per Ni basis) decreases with $x$. A rapid drop from $x = 1.0$ to $0.9$ becomes more gradual and linear, with an extrapolation to $\Delta S_{mag} = 0$ at $x = 0.4$, coincident with the $T = 0$ phase boundary between stripe insulator and correlated metal. Finally, the Sommerfeld plot in Figure 3(c) shows a clear separatrix between gapped specimens ($x \geq 0.5$) and metals ($x \leq 0.4$). We attribute the small, finite $\gamma$ found at $x = 0.5$ to the same degree of inevitable chemical heterogeneity influencing the transport behavior of samples near the phase boundary (*vide supra*).

**Single Crystal Diffraction.** As we have shown for $La_4Ni_3O_8$,[19] the charge stripe phase can be identified using X-ray scattering by the appearance of in plane superlattice reflections indexed on a propagation vector $\mathbf{q} = (1/3,1/3)$ in the reciprocal basal plane of the $I4/mmm$ setting. Figure 4 shows that $(Pr_{1-y}La_x)_4Ni_3O_8$ crystals with $x \geq 0.5$ exhibit such a superlattice below $T_{MIT}$, implying a similar stripe structure for these substituted specimens. The $(h,k,0)$ scattering plane for an $x = 0.7$ specimen is shown in Figures 4a and b for $T < T_{MIT}$ and $T > T_{MIT}$, respectively. In plane superlattice reflections at $\mathbf{q} = (1/3,1/3)$ are observed in the low-temperature data, which disappear upon warming through the insulator-metal transition. An analysis of the scattering intensity reveals that the charge stripe stacking arrangements both within the plane and normal to it are unchanged from that found for the $La_4Ni_3O_8$ endmember.[19] That is, stripes are stacked on top of one another within a trilayer and follow an ABAB pattern between trilayers (see Ref. 19). Interestingly, the position and intensity distribution of the superlattice reflections both in the $a^*b^*$-plane and along $c^*$ is largely independent of composition, as shown in the line cuts through superlattice reflections in Figure 4c,d. That is, the charge stripe phase remains largely unchanged upon Pr substitution for La for x > 0.4. This behavior is not unreasonable because $Pr^{3+}$ and $La^{3+}$ are isovalent; the electron concentration that sets the stripe periodicity is unchanged with $x$. As expected, the peak width along $c^*$ is significantly broader than that for in-plane reflections, a result of weak inter-trilayer coupling. Indeed, for $La_4Ni_3O_8$, the correlation length $x \approx 8$ Å along $c$.[19] Similar behavior appears to be the case here. In the plane, we do not find evidence for finite-size broadening of the superlattice peaks in these data across the series. Thus, within our experimental resolution, the charge stripe phase does not lose coherence as the phase boundary at $x = 0.4$ is approached from above. It is possible that such loss of coherence may appear if the composition were closer to the boundary than our current sampling interval allows.

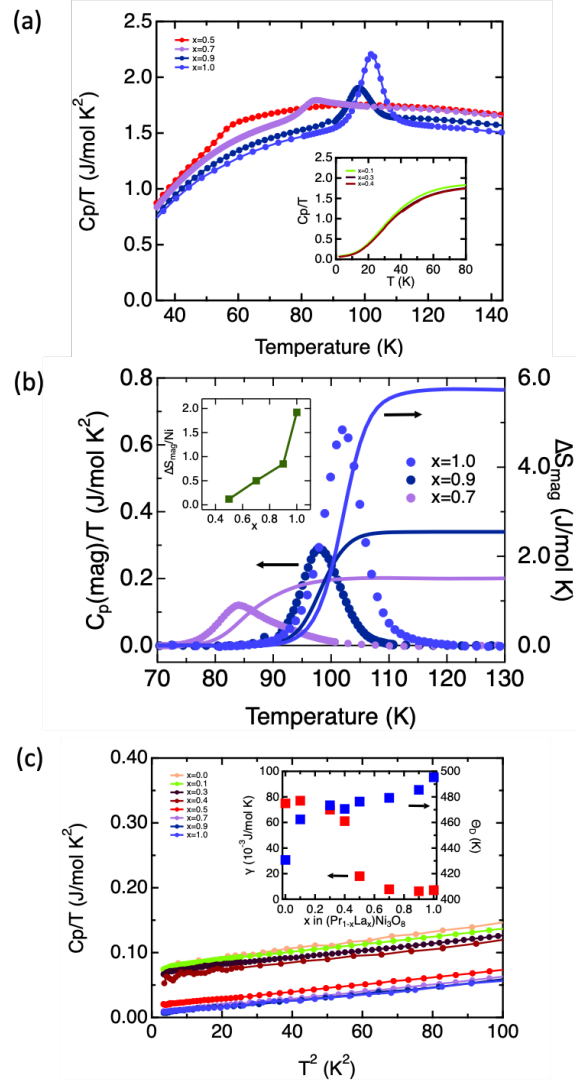

**Figure 3**. (a) Heat capacity of $(Pr_{1-x}La_x)_4Ni_3O_8$ ($x$ = 0.5, 0.7, 0.9, 1.0) single crystals. Data for $x$ = 1.0 from Ref. 19. Inset: $x$ = 0.1, 0.3, 0.4. (b) Magnetic heat capacity (left) and associated magnetic entropy (right) for $x$ = 0.7, 0.9, 1.0. Inset: $\Delta S_{mag}$/Ni vs composition. (c) $C_p/T$ vs $T^2$. Inset: Sommerfeld coefficient (left) and Debye temperature (right) vs. composition. See text for details.

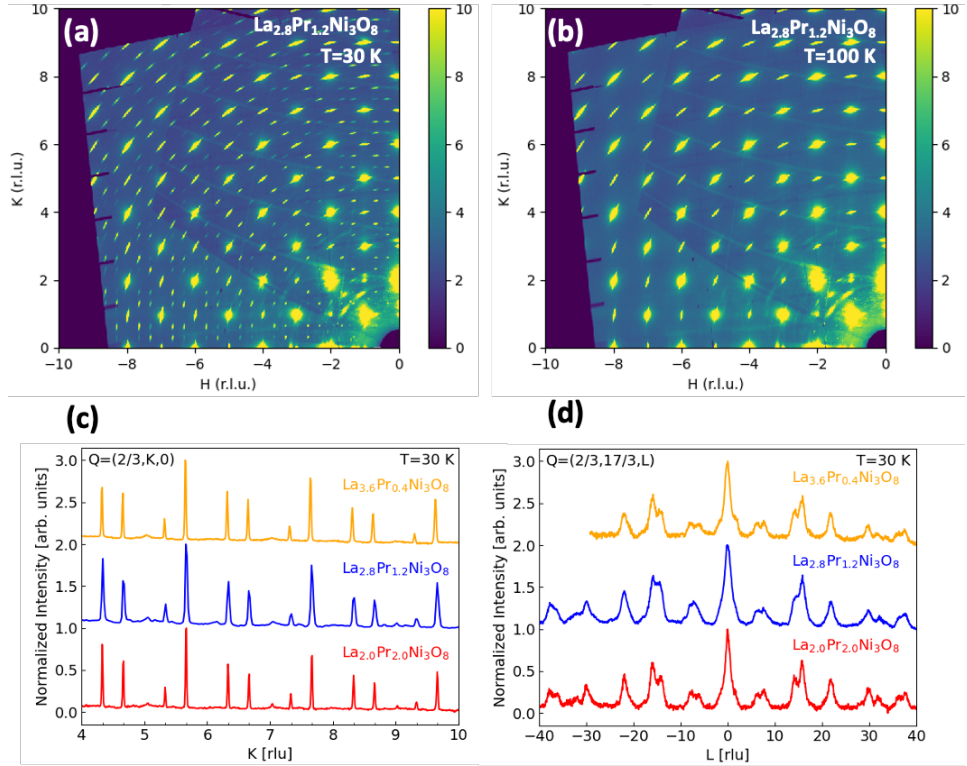

**Figure 4**. Single crystal X-ray diffraction. $(h,k,0)$ scattering plane for $(Pr_{0.3}La_{0.7})_4Ni_3O_8$ ($x = 0.7$) at (a) T = 30 K and (b) 100 K. Line cuts at 30 K along (c) $(2/3, k, 0)$ and (d) $(2/3, 17/3, l)$ for samples with $x = 0.5, 0.7, 0.9$. In (c) and (d) curves have been offset for clarity.

**DISCUSSION**

Cumulatively, the data presented above allow us to propose an electronic and magnetic phase diagram for the $(Pr_{1-x}La_x)_4Ni_3O_8$ system, which is shown in Figure 5a. Most of the phase diagram is occupied by a correlated metal phase at high temperature, and as the ground state for $x < 0.4$. For compositions $x > 0.4$, the charge-stripe phase is found below the solid line. Other features observed in the magnetic data are identified on the diagram as well. First is the locus of the broad maximum observed in the $c$-axis magnetization for $x \leq 0.5$, with a speculated extension into the stripe phase region. The fact that this maximum occurs more prominently as the Pr content increases and that it is absent for $\boldsymbol{H} \parallel \boldsymbol{ab}$ suggests that it could reflect thermal population of crystal field levels on the $Pr^{3+}$ ions. Other more exotic possibilities, such as precursory spin-stripe order that remains short-range for $x < 0.5$ but that coarsens when the charge stripes form at $x \geq 0.5$, should also be considered. Second, a nearly composition-independent line is found at $T \approx 30$ K, tracking a local minimum in the $c$-axis magnetization. Again, this feature weakens as Pr is diluted by La, and we attribute the low temperature upturn that gives rise to this minimum to Pr-localized spins (See detailed discussion in the SI).

We now present two plausible models for the phase evolution across the solid solution, beginning with a picture in which Pr addition to the charge- and spin-stripe phase in $La_4Ni_3O_8$ ($x = 1$) nucleates an inhomogeneous mixture of La-rich stripe phase and Pr-rich regions that evolves into a uniform metallic phase below $x \approx 0.4$. A schematic of this evolution is shown in Figure 5b. As Pr is lightly substituted into La$_4$Ni$_3$O$_8$, isolated Pr-rich regions will be formed statistically. Since the pure Pr$_4$Ni$_3$O$_8$ endmember shows no long-range charge- or spin-stripe order,[20] we conjecture that in these regions the stripe order will be quenched. As more Pr substitutes for La, the volume of stripe phase will decrease, albeit less efficiently as the Pr 'spheres of influence' begin to overlap. Eventually, the metallic regions percolate at $x = 0.4$.

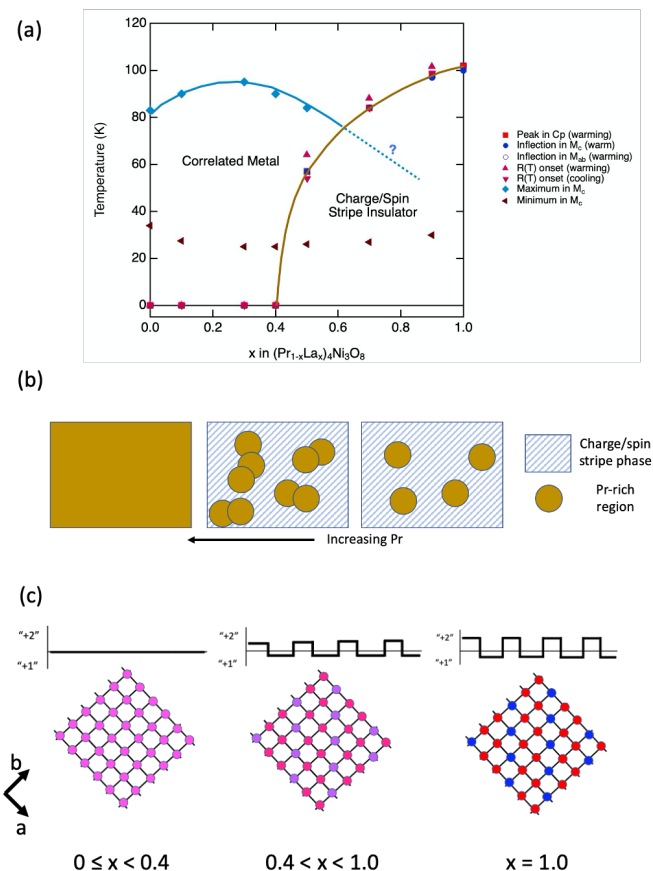

**Figure 5**. (a) Electronic and magnetic phase diagram of (Pr$_{1-x}$La$_x$)$_4$Ni$_3$O$_8$ from single crystals. (b) Proposed schematic model in which an inhomogeneous mixture of charge- and spin-stripe ordered regions evolves with increasing Pr content into a uniform metallic phase below $x = 0.4$. (c) Alternative model in which the amplitude of the charge- and spin stripes continuously decreases away from the $x = 1$ endmember. Red and blue circles represent Ni in formally 1+ and 2+ states at $x = 1$. The charge separation becomes less pronounced for $0.4 < x \leq 1.0$, shown by the lighter blue and red circles, and finally, purple circles denote a uniform "1.33+" Ni state with no stripes in the metallic regime. The continuously decreasing amplitude of the charge stripe is shown schematically at the top. Stripes are oriented at 45° from the crystal axes, see Ref. 19.

Assuming the suppression of magnetic entropy with decreasing $x$ measures a phase volume effect, then the mixed-phase picture agrees with data of Figure 3b, in which a rapid drop in $\Delta S_{mag}$ as $x$ decreases initially from $x = 1.0$ is followed by a more gradual but linear decrease of $\Delta S_{mag}$. The phase boundary occurs where the Pr site fraction is 0.6, remarkably close to the 2D square lattice critical site fraction for percolation ($p_c = 0.59$). However, it is unclear if substitution of Pr for La in the fluorite layer would be as effective at disrupting the stripe order as substitution within the nickel oxide trilayer blocks (see Figure 1 a), potentially impacting the observed percolation concentration. Furthermore, it is not obvious within this mixed-phase picture why $\Delta S_{mag}$ should extrapolate continuously to zero at $x = 0.4$ (Figure 3b) rather than to a finite value. Finally, it is difficult to understand the $x$-dependence of $T_{MIT}$ without invoking some unknown influence of the Pr-rich regions that would change the interactions in the ordered stripe phase that set $T_{MIT}$.

In an alternative model, the system remains a single, homogeneous charge- and stripe-ordered phase for $x > 0.4$, but with addition of Pr the amplitude of the charge and spin disproportionation continuously decreases, vanishing at the phase boundary into a uniform metallic phase. In such a scenario, schematically diagrammed in Figure 5c, the decrease in $\Delta S_{mag}$ results directly from the reduced spin amplitude at the ordering transition. We would expect this second-order-like process to show a continuous decrease in $\Delta S_{mag}$ to zero as the phase boundary at $x$=0.4 is approached, a behavior observed in Figure 3b. Additionally, the variation of $T_{MIT}$ with $x$ is more naturally understood in this model, since the interactions setting the transition energy scale should depend on the amplitude of the charge disproportionation.

Unfortunately, the X-ray diffraction results presented above do not help us to distinguish between a homogeneous or inhomogeneous state below $T_{MIT}$. First, because $La^{3+}$ and $Pr^{3+}$ are isovalent, the electron concentration that sets the periodicity of the stripe phase is expected to remain fixed at 1/3 independent of $x$ whether the system is electronically homogeneous or inhomogeneous. Second, the intensity of the superlattice reflections is proportional to a product of the volume occupied by the ordered phase and the square of its (unknown) structure factor. With only the single X-ray measurement, these two factors cannot be deconvoluted.

We finish by returning to the issue of the nature of the $T = 0$ phase transition between the correlated metal for $x < 0.4$ and the charge- and spin-stripe insulator for $x > 0.4$. If the mixed-phase model is appropriate, then the presence of disorder introduced by the Pr/La substitution situates the $(Pr_{1-x}La_x)_4Ni_3O_8$ system as a candidate quantum Griffiths phase, with expected variation of critical exponents as a function of both the control parameter, $x$, and temperature that would be absent in a clean system.[26] Alternatively, if the homogeneous scenario is correct, then the expected continuous evolution of the charge- and stripe-order parameter toward $x = 0.4$ implies the possibility of a clean quantum critical point separating the correlated metal and stripe phase insulator.

The study of such states of matter lies outside the scope of this work. Nonetheless, our results show the value of the $(Pr_{1-x}La_x)_4Ni_3O_8$ system for exploring either of these two alternative quantum phase transitions and offer two possible materials pathways forward to better characterize the phenomenology of the boundary region. To better understand the behavior of the phase boundary, samples could be grown with compositions closer to the putative quantum phase transition at $x = 0.4$. While technically possible, inevitable composition fluctuations such as those shown in the electrical transport data (Figure 2a), will complicate the interpretation. A more promising route would be to use pressure as a control parameter. As shown in Figure 1c, $dV/dx$ is positive. It may thus be possible to apply pressure to a specimen whose composition is sufficiently close to, but slightly above, $x = 0.4$ to drive the system toward the quantum phase transition without introducing additional disorder. This approach has the additional benefit of allowing the analog tuning of the pressure control parameter, which would be impossible in a 'digital' compositional study. We note that in his own exploration of the MIT in polycrystalline $La_4Ni_3O_8$,[31] Goodenough used pressure to explore the suppression of the MIT. It is fitting that a decade later we see his insights into the control of charge and spin states as the pathway to understanding this larger family of complex oxides.

**CONCLUSIONS**

Using single crystals, we have presented a comprehensive study of $(Pr_{1-x}La_x)_4Ni_3O_8$, including an electronic and magnetic phase diagram of this system. Specifically, we have identified the approximate location of a quantum phase transition at $x \approx 0.4$, where metals are found at $x < 0.4$ and a charge- and spin-stripe phase is found at $x > 0.4$. Importantly, the charge-stripe phase appears to be much the same as that found in the pure $La_4Ni_3O_8$ endmember. We suggest two possible models, heterogeneous and homogeneous, to explain the $x$-dependence of the competing phases, leading to the possibility of either potential quantum Griffiths phase physics or of a quantum critical point at $x \approx 0.4$. In either case, our work establishes the relevance of the $(Pr_{1-x}La_x)_4Ni_3O_8$ system for future exploration of these $T = 0$ phenomena.

## ASSOCIATED CONTENT

**Supporting Information**

The Supporting Information is available free of charge on the ACS Publications website at DOI:.

In-plane Magnetization; Extrinsic vs. intrinsic behavior; Low temperature magnetic behavior. (PDF)

## AUTHOR INFORMATION

**Corresponding Authors**

*E-mail: mitchell@anl.gov (J.F.M.).

*E-mail: xinglong.chen@anl.gov (X.C.).

**Notes**

The authors declare no competing financial interest.

## ACKNOWLEDGMENTS

This work was supported by the U.S. Department of Energy, Office of Science, Office of Basic Energy Sciences, Materials Science and Engineering Division.  Use of the Advanced Photon Source was supported by U.S. Department of Energy, Office of Science, Office of Basic Energy Sciences under Contract Number No. DE-AC02-06CH11357.

# Supporting Information

# Competing Charge/Spin-Stripe and Correlated Metal Phases in Trilayer Nickelates $(Pr_{1-x}La_x)_4Ni_3O_8$

Xinglong Chen,[1,*] Hong Zheng,[1] Daniel Phelan,[1] Hao Zheng,[1] Saul Lapidus,[2] Matthew J. Krogstad,[2] Raymond Osborn,[1] Stephan Rosenkranz,[1] and J. F. Mitchell[1,*]

[1]Materials Science Division, Argonne National Laboratory, Lemont, Illinois 60439, United States

[2]Advanced Photon Source, Argonne National Laboratory, Lemont, Illinois 60439, United States

*To whom correspondence should be addressed

E-mail: E-mail: mitchell@anl.gov (J.F.M.); xinglong.chen@anl.gov (X.C.)

Contents:

## I.    In-plane Magnetization

Figure S1 shows the in-plane magnetization of $(Pr_{1-x}La_x)_4Ni_3O_8$ specimens as a function of temperature. Although weaker than that seen in the *c*-axis data (Figure 2, main text), the onset of spin order (stripes) is seen for $x \geq 0.5$ (arrows). The anisotropy between *ab*-plane and *c*-axis magnetization is clear, including the absence of the broad maximum near 80–100 K, the minimum near 30 K, and the low temperature upturn, all features found for $H \parallel c$.

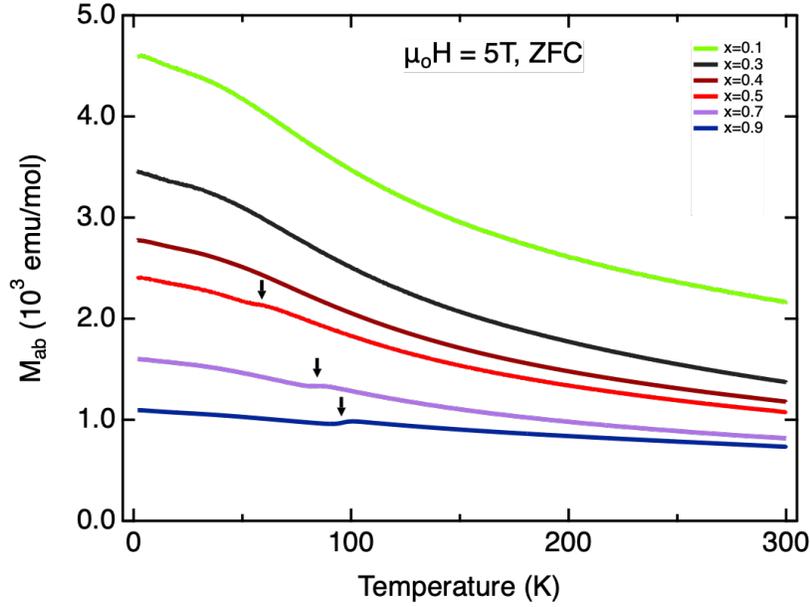

## II.    Extrinsic vs. Intrinsic Behavior

As previously reported by numerous authors studying reduced nickelates containing $Ni^{1+}$, a (super)paramagnetic component is found and attributed to 'nickel nanoparticles' that are a potential byproduct of synthesis by hydrogen or hydride reduction. In contrast, Huangfu *et al.* have argued that this behavior is intrinsic to powders of $Pr_4Ni_3O_8$ and not reflective of extrinsic Ni agglomerates.[S1,S2] [REFS] We have investigated this behavior in single crystals of the $(Pr_{1-x}La_x)_4Ni_3O_8$ system where we can access anisotropic characteristics of the magnetization to see if further insight is possible into this question.

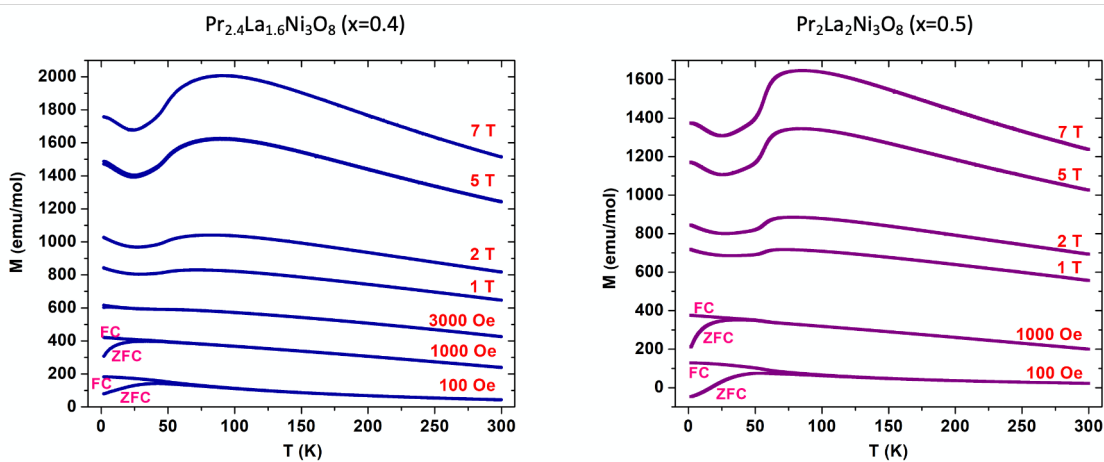

**Figure S2**. Zero-field cooled (ZFC) and field-cooled (FC) magnetization vs. temperature for an $x$=0.4 single crystal (left) and $x$=0.5 single crystal (right) with $\boldsymbol{H} \parallel \boldsymbol{c}$.

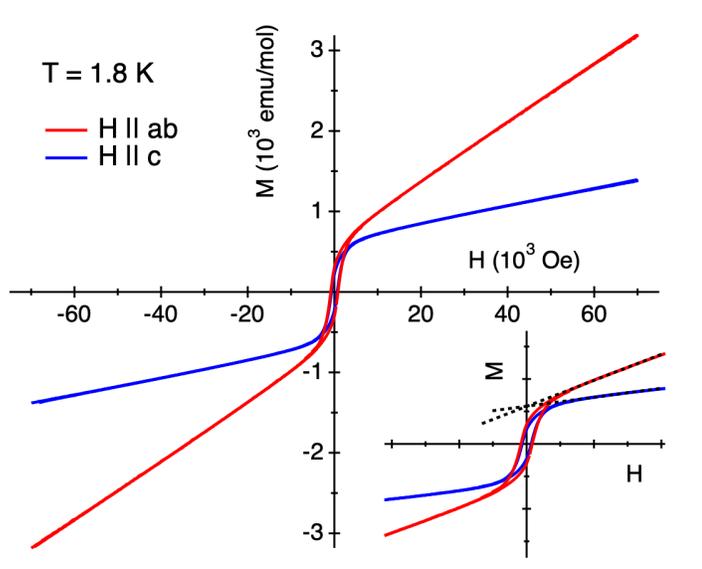

Figure S2 shows $M(T)$ for $x = 0.4$ and $x = 0.5$ specimens measured in ZFC and FC modes using fields from 100 Oe to 5 T directed along $\boldsymbol{c}$. In low fields, an irreversibility extending to near 50 K is observed, but this bifurcation between ZFC and FC curves vanishes above 1 T, suggesting that a ferromagnetic-like component has become saturated. That this is the case can more clearly be seen in Figure S3, where $M(H)$

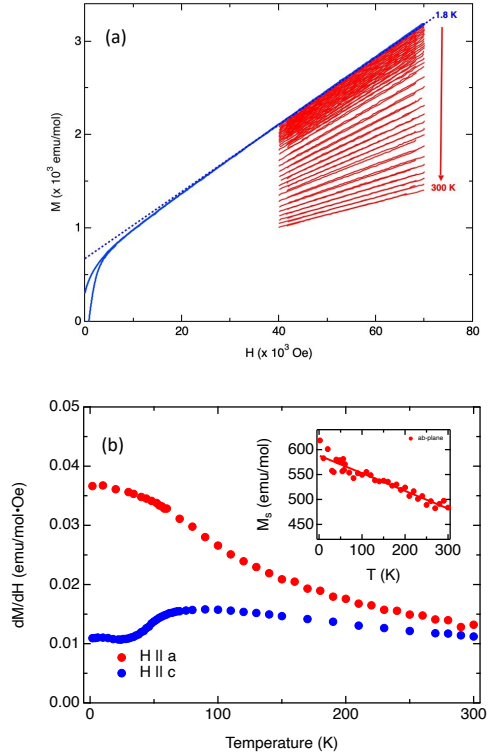

**Figure S4.** (a) $M(H)$ for $Pr_2La_2Ni_3O_8$ ($x$=0.5) in the range 4-7 T, 1.8 K < T < 300 K. $\boldsymbol{H} \parallel \boldsymbol{ab}$. (b) Intrinsic susceptibility of $Pr_2La_2Ni_3O_8$ evaluated as $dM/dH$ in the range 4-7 T. Inset: saturation magnetization, $M_s$, extracted at 1.8 K is plotted for $\boldsymbol{H} \parallel \boldsymbol{ab}$ and $\boldsymbol{H} \parallel \boldsymbol{c}$. A small but clear hysteresis is observed in both channels, with a coercivity of roughly 800 Oe, explaining the loss of ZFC-FC irreversibility at and above 1 T. Also clear in Figure S3 is a pronounced anisotropy.

Above approximately 2 T, $M$ vs $H$ at T = 1.8 K is strictly linear. This same linear $M(H)$ is found in the entire temperature range to 300 K, as shown in Figure S4a. Huangfu et al. report this behavior as well in $Pr_4Ni_3O_8$ powders.[S3] Following the usual approach used by those authors for data such as these, and writing the total magnetization, $M_{tot}$, as the sum of two components—$M_0$, a saturating ferromagnetic phase, and $M_1$, a nonsaturating phase, the magnetic susceptibility of the remainder of the sample is simply given by the slope of this line, while the intercept with the $H = 0$ axis measures the full magnetization of the saturating ferromagnetic component. This derived susceptibility is plotted in Figure S4b for $\boldsymbol{H} \parallel \boldsymbol{ab}$ and $\boldsymbol{H} \parallel \boldsymbol{c}$, showing the emergence of anisotropy below 300 K and growing to nearly a factor of four at 1.8 K. Note also the distinct qualitative behavior in these two directions, with the $c$-axis susceptibility showing the broad maximum and a low T upturn not found in the $\boldsymbol{ab}$-plane susceptibility. The data are quantitatively like those shown in Fig. 2 of Ref. S3, albeit these data are powder averaged and for a different composition, i.e., $Pr_4Ni_3O_8$ ($x = 0$ in our notation). Fitting the powder averaged single crystal susceptibility $\chi_{avg} = (1/3)(2\chi_{ab} + \chi_c)$ to a Curie-Weiss law in the range 60–300 K (i.e. above the stripe ordering temperature for $x = 0.5$) yields a Curie constant of 5.8 emu K mol⁻¹ Oe⁻¹ and $\theta = 155$ K, in reasonable agreement with that reported by Huangfu et al.,[S3] (6.8 emu K mol⁻¹ Oe⁻¹ and 160 K, respectively), albeit on $Pr_4Ni_3O_8$.

The inset to Figure S4b shows the temperature evolution of the intercept of a linear fit to the $ab$-plane $M(H)$ for $Pr_2La_2Ni_3O_8$ ($x = 0.5$), attributed to the magnetization of the saturating component. Notably, this remains large even at 300 K. Although we are unable to fit our data, which appears linear up to 300 K, to the 1-$(T/T^*)^{3/2}$ mean-field form used in Ref. S3, the value at 1.8 K is 600 emu/mol, or 0.1 $\mu_B$/f.u.-Ni, which is comparable to the 0.22 $\mu_B$/f.u. reported by Huangfu et al.[S3] for unsubstituted $Pr_4Ni_3O_8$. There, it was claimed that the size of this moment would imply $\approx 10\%$ of the original $Pr_4Ni_3O_{10}$ would have had to decompose into Ni nanoparticles, an amount detectable by X-ray diffraction. In the absence of such evidence, they conclude the behavior is intrinsic to $Pr_4Ni_3O_8$.

Thus far, our data are largely consistent with those of Huangfu et al., which raises the question of whether our single crystal data also imply an intrinsic origin to the ferromagnetic-like response. As we now discuss, the field-dependent behavior found in our single crystals is difficult to reconcile within the intrinsic picture. The inset of Figure S3 shows $M(H)$ along the $ab$-plane and $c$-axis, with extrapolations of the linear fit from high field. First, we note that the coercive field is nearly identical for both directions. Second, the intercept at $H = 0$ is the same for both $H \parallel ab$ and $H \parallel c$. Both points argue that the ferromagnetic component at low temperature is *isotropic*. Considering that we see a clear anisotropy in all other measured properties of the single crystals, and in particular a pronounced magnetization anisotropy below 300 K (Figure S4b), it is difficult to understand how this ferromagnetic-like property would not also exhibit anisotropy if it were intrinsic to the trilayer nickelate. More work is needed to clarify the origin of this behavior, which appears to be common across sample type and groups, and has now been characterized in infinite layer nickelates as well.[S2]

### III.     Low Temperature Magnetic Behavior

As discussed in the main text, the $c$-axis (but not $ab$-plane) magnetization exhibits a low temperature upturn. Figure S5 shows a plot of the size of this upturn, phenomenologically defined as the height, $h$, of the curve at T = 1.8 K with respect to the minimum near 30 K in all specimens. As shown in Figure S5, $h$ varies linearly with $x$, extrapolating to $h = 0$ at $x = 1$, the $La_4Ni_3O_8$ endmember. This behavior leads us to assign the upturn not to the nickel sublattice, but rather to Pr spins.

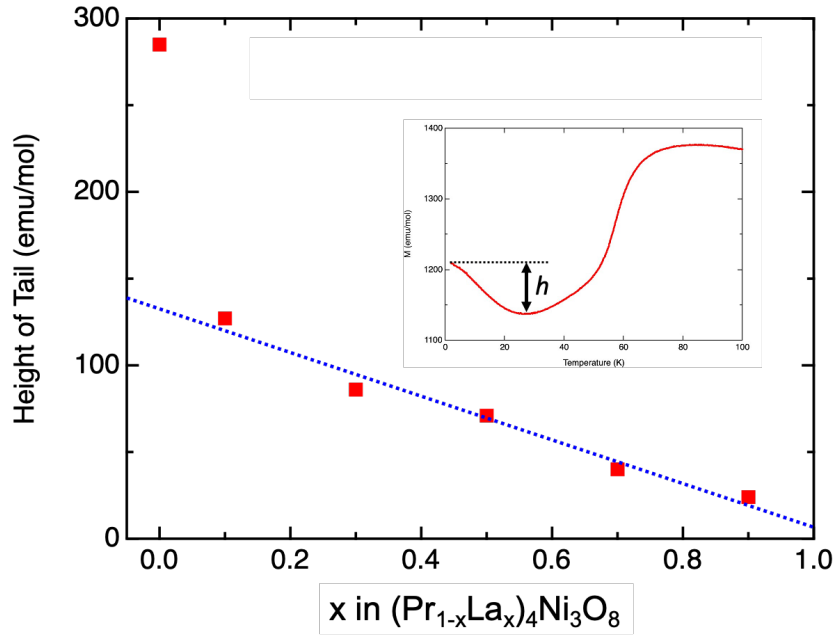

**Figure S5**. Height of *c*-axis magnetization upturn vs. composition. Inset: definition of height parameter, *h*.